\documentclass[final,3p,times]{elsarticle}

\usepackage[numbers]{natbib}
\usepackage{latexsym}
\usepackage{amssymb}
\usepackage[mathscr]{eucal}
\usepackage{epsfig}
\usepackage{lscape}
\usepackage{amssymb}
\usepackage{amsmath}
\def\Bfg{\boldsymbol}

\journal{Applied Numerical Mathematics}

\def\e{\mathop{\rm e}\nolimits}

\def\d{{\rm d}}
\def\ud{{\textstyle{\frac{1}{2}}}}

\def\Green{\relax}
\newskip\tableskipamount \tableskipamount=8pt plus 3pt minus 3pt
\def\tableskip{\vskip\tableskipamount}
\def\fileth{\noalign{\hrule}}
\def\g{\tilde g}

\begin{document}

\begin{frontmatter}

\title{Summation of divergent series: Order-depen\-dent mapping}

\author{Jean Zinn-Justin}
\ead{jean.zinn-justin@cea.fr} 
\address{CEA, IRFU and Institut de Physique Th\'eorique,
Centre de Saclay,  F91191 Gif-sur-Yvette Cedex, France}

%

\begin{abstract}


Summation methods play a very important role in quantum field theory  because all perturbation series are divergent and the expansion parameter is not always small. A number of methods have been tried in this context, most notably Pad\'e approximants, Borel--Pad\'e summation, Borel transformation with mapping, which we  briefly describe and one on which we concentrate here,  Order-Dependent Mapping (ODM). We recall the basis of the method, for a class of series we give intuitive arguments to explain its convergence and illustrate its properties by several simple examples. Since the method was proposed, some rigorous convergence proofs were given. The method has also found  a number of applications and we shall list a few.

\end{abstract}
\end{frontmatter}
\section{The initial motivation: Perturbative quantum field theory}

In quantum field theory, the main analytic calculation tool is the perturbative expansion.  As an illustration, we consider the important example of the \Green{$\phi^4$} field theory \cite{ZJbook}. In the statistical formulation, one considers the Euclidean (or imaginary time) action 
\Green{$\mathcal{  S}$}, local functional of the field \Green{$\phi(x)$, $x\in\mathbb{ R}^d$},
\begin{equation} 
 \mathcal{ S}(\phi)=\int\d^d x\,\left[\ud \sum_{\mu=1}^d\left[\partial_\mu\phi(x)\right]^2+\ud r\phi^2(x)+\frac{g}{4!}\phi^4(x)\right], \label{eODMfiv} 
 \end{equation}
 where $r$ and $g$ are two parameters.
To this action is associated a functional measure $\e^{-\mathcal{ S}(\phi)}/\mathcal{ Z}$, where $\mathcal{ Z}$ is the partition function given by the field integral
\begin{equation}
\mathcal{ Z}=\int[\d\phi]\e^{-\mathcal{ S}(\phi)}.\label{eODMZfiv}
 \end{equation}
The limit \Green{$d=0$} corresponds to a simple integral.\par
The case \Green{$d=1$} corresponds to the quantum quartic anharmonic oscillator. 
\par
Dimensions $d>1$  correspond to quantum field theory and the expression (\ref{eODMfiv}) is then somewhat symbolic since the theory has to be modified at short distance to regularize UV divergences and renormalized to cancel them.\par
In particular, the dimensions \Green{$d=2,3$} are especially relevant to classical statistical physics and the theory of phase transitions. Finally, \Green{$d=4$} is relevant to the theory of fundamental interactions at the microscopic scale. The corresponding relativistic quantum field theory is part of the so-called Higgs mechanism.\par
For the field theory (\ref{eODMfiv}), the perturbative expansion  amounts to an expansion in powers of the positive parameter \Green{$g$}. \par
For \Green{$d>1$}, the difficulty of evaluating the successive  perturbative terms increases very rapidly. Moreover, questions like regularization and renormalization arise. Therefore, the calculation of   renormalization group functions in the \Green{$d=3$} \Green{$(\Bfg{\phi}^2)^2$} field theory \textit{up order $g^7$} \cite{rBNGMo} \textit{is a remarkable achievement.} 

\subsection{Large order behaviour of perturbative series}

In the $\phi^4$ field theory (\ref{eODMfiv}), $g=0$ corresponds to a singularity since the integral (\ref{eODMZfiv}) is not defined for $g<0$. \label{LOBphiiv} The \textit{perturbative series is divergent.} 
For $d<4$, the large order behaviour  can be inferred from a steepest descent calculation of the field integral (\ref{eODMZfiv}) \cite{rLOBLip} \cite{rLOBgen}. For the quartic anharmonic oscillator ($d=1$) the result was derived earlier from the Schrödinger equation \cite{rCBTTW}.  For any physical observable $f$, the results have the general structure
\begin{equation}
f_k \mathop{\propto}_{k\to\infty} (-1)^k k^b a^k k!\,,\label{eLObehaviour}
\end{equation}
where $a$ depends only on $d$ and $b$ is a half-integer that depends on the observable. The coefficient $A=1/a$ has the value  
\begin{eqnarray} 
d=0:\ &A&=3/2\,, \\
d=1:\   &A&=8\quad  \,,  \\
d=2:\   &A& =35.10268957367896(1)\quad  \textrm{(Zinn-Justin)}, \\
d=3:\    &A& = 113.38350781527714(1)\quad  \textrm{(Zinn-Justin)}\label{eLOBfivDiii}.  
\end{eqnarray}
For \Green{$d=4$}, to the contribution coming from the steepest descent calculation, a contribution due to the large momentum singularities of Feynman diagrams has in general to be added.\par
 Finally, notice that for $d<4$, Borel summability has been proved.
\par
Similar results can be obtained for a number of quantum field theories. When the formal expansion parameter is Planck's constant, a divergence of the form (\ref{eLObehaviour}) is in general found (except for some fermion theories), but the parameter $a$ may be complex. For an early review, see \cite{ZJLOreport}.
\par
 It follows from the large order behaviour analysis that, when the expansion parameter is not small, a summation of the perturbative expansion is indispensable.

\subsection{Series summation}

In the study of the fundamental interactions at the microscopic scale, it was realized that  in the case of the strong nuclear force, unlike QED, the expansion parameter was large and, therefore, perturbation theory useless, leading many physicists even to reject quantum field theory as a framework to describe such phenomena.\par
Before the large order behaviour was even known, in \cite{rDBMP} it was proposed, instead, to sum the perturbative expansion, using Pad\'e approximants and the idea was applied to a phenomenological model, the $\phi^4$ field theory in $d=4$ dimensions. 
 Since only two or three terms could be calculated, the possible convergence of the Pad\'e summation could not be checked very well. However, the results obtained in this way made much better physical sense than those of plain perturbation theory. For a review see \cite{ZJthesis}.

\par
In the seventies, one outstanding problem for which summation methods was required, is the determination of critical exponents and other critical quantities in the theory of second order phase transitions. Following Wilson, for a whole class of physical systems, these quantities can be obtained from the $({\Bfg \phi}^2)^2$ field theory in $d=3$ dimensions. One verifies immediately that the expansion parameter, the renormalized interaction $g_{\rm r}$, is of order $1$ and a series summation is required (we do not discuss here the $\varepsilon=4-d$ expansion, but the problem is analogous).\par

To deal with the practical problem of series summation, a method was proposed based on   Borel--Pad\'e approximants \cite{rBNGMo}. With the knowledge of the large order behaviour, a more efficient method could be developed, combining a Borel transformation (actually Borel--Leroy) and a conformal mapping  \cite{rLGZJ}, \cite{rRGZJ}, which we briefly present in next section. However, another method based only on the analytic properties of the series, the order-dependent mapping   was also investigated, which we describe in more detail in section \ref{ssODMdef} (a general reference is \cite{ZJbook}).

\begin{table}[t]
$$  \offinterlineskip\tabskip=0pt\halign to \hsize
{& \vrule#\tabskip=0em plus1em & \strut\ # \ 
& \vrule#& \strut #
& \vrule#& \strut #  
& \vrule#& \strut #  
& \vrule#& \strut #  
& \vrule#& \strut #  
& \vrule#& \strut # 
&\vrule#\tabskip=0pt\cr
\noalign{ \caption{} \label{tabCPii}\tableskip}
\noalign{\noindent\it Series  summed by the method based on Borel
transformation and mapping for the zero $\g^* $ of the RG $\beta(g) $ function
and the exponents $\gamma$ and $\nu$ in the $ \phi^4_3 $ field
theory.\tableskip} 
\fileth
height2.0pt& \omit&& \omit&& \omit&& \omit&& \omit&& \omit&& \omit&\cr
&$ \hfill k \hfill$&&$ \hfill 2 \hfill$&&$ \hfill 3 \hfill$&&$ \hfill 4
\hfill$&&$ \hfill 5 \hfill$&&$ \hfill 6 \hfill$&&$ \hfill 7 \hfill$&\cr
height2.0pt& \omit&& \omit&& \omit&& \omit&& \omit&& \omit&& \omit&\cr
\fileth
height2.0pt& \omit&& \omit&& \omit&& \omit&& \omit&& \omit&& \omit&\cr
&$ \hfill\g^*  \hfill$&&$ \hfill 1.8774  \hfill$&&$ \hfill 1.5135 \hfill$&&$
\hfill  1.4149 \hfill$&&$ \hfill  1.4107 \hfill$&&$ \hfill 1.4103 \hfill$&&$
\hfill 1.4105 \hfill$&\cr 
height2.0pt& \omit&& \omit&& \omit&& \omit&& \omit&& \omit&& \omit&\cr
\fileth
height2.0pt& \omit&& \omit&& \omit&& \omit&& \omit&& \omit&& \omit&\cr
&$ \hfill \nu \hfill$&&$ \hfill 0.6338  \hfill$&&$ \hfill 0.6328 \hfill$&&$
\hfill 0.62966 \hfill$&&$ \hfill 0.6302 \hfill$&&$ \hfill 0.6302 \hfill$&&$
\hfill0.6302  \hfill$&\cr 
height2.0pt& \omit&& \omit&& \omit&& \omit&& \omit&& \omit&& \omit&\cr
\fileth
height2.0pt& \omit&& \omit&& \omit&& \omit&& \omit&& \omit&& \omit&\cr
&$ \hfill \gamma \hfill$&&$ \hfill 1.2257 \hfill$&&$ \hfill  1.2370
\hfill$&&$ \hfill   1.2386\hfill$&&$ \hfill 1.2398 \hfill$&&$ \hfill 1.2398
\hfill$&&$ \hfill  1.2398 \hfill$&\cr 
height2.0pt& \omit&& \omit&& \omit&& \omit&& \omit&& \omit&& \omit&\cr
\fileth} 
$$
\end{table}
\section{Borel transformation and conformal mapping}

The values of critical exponents in a large class of continuous (or second order) phase transitions can be inferred from so-called renormalization group (RG) functions of the $(\Bfg{\phi}^2)^2$ quantum field theory.  \label{ssBorelmapping} One important function is the RG $\beta$-function
whose zeros determine the RG fixed points. For example, in the case of the $\phi^4$ theory in $d=3$ dimensions, 
Nickel \cite{rBNGMo} has calculated
\begin{eqnarray}
\tilde\beta(\g)  & =&-\g+ \g^2 -\frac{308}{729} \g^3 + 0.3510695977\g^4
   -0.3765268283 \g^5    + 0.49554751 \g^6 \nonumber \\  &\quad&  - 0.749689 \g^7
+O\left(\g^8\right) ,\label{ebetafiviiiIsing}   
\end{eqnarray}
where  $\g =3g_{\rm r}/(16\pi) $ and $g_{\rm r}$ is the so-called renormalized interaction, related to the parameter that appears in the action (\ref{eODMfiv}) by $g_{\rm r}=g+O(g^2)$ and
\begin{equation}
\beta(g_{\rm r})= \frac{16\pi}{3}\tilde\beta(\g)=-\frac{1}{\d\ln g/\d g_{\rm r}}.\label{erbetadef} 
\end{equation}
The perturbative expansion is \textit{divergent} (equation (\ref{eLObehaviour})). For the \Green{$\beta$}-function in three dimensions, 
$$\tilde\beta(\g)=\sum_k \tilde\beta_k\, \g^k,$$
the large order behaviour, implied by the estimate (\ref{eLOBfivDiii}), is given by  
\Green{$$  \tilde\beta_k\mathop{\propto}_{k\to\infty}(-a)^k k^{7/2} k!  $$}
with \Green{$a  =0.147774232\ldots$}. 
\par
To characterize the large distance properties of statistical systems at the phase transition, one must first determine the non-trivial zero $\g^*$ of the \Green{$\beta $}-function and then calculate various physical quantities like critical exponents for $\g=\g^*$. One discovers that $\g^*$ is a number of order 1 and, thus, a numerical determination from the series \ref{ebetafiviiiIsing} clearly requires a summation of the series. \par
In three dimensions, the perturbative expansion is proved to be \textit{Borel summable}. It is thus natural to introduce the Borel--Laplace transformation (here, Borel--Leroy): 
$$B_\sigma(g)=\sum_k \frac{\beta_k}{\Gamma(k+\sigma+1)}g^k,$$
where $\sigma $ is a free parameter.
 Then, formally in the sense of power series 
\Green{$$\beta(g)=\int_0^{+\infty}t^\sigma\e^{-t}B_\sigma(gt)\d t\,.$$}
The function \Green{$B_\sigma(g)$} is analytic in a circle of radius \Green{$1/a$}. The series is said Borel summable if, in addition, \Green{$B_\sigma(g)$} is analytic in a neighbourhood of the real positive semi-axis and the integral converges. 
\par
The series defines the function in a circle. It is thus necessary to perform an analytic continuation. In practice, with a small number of terms, the continuation requires a  domain of analyticity larger than rigorously established.
Le Guillou and Zinn-Justin \cite{rLGZJ} have assumed maximal analyticity, \textit{i.e.},
analyticity in a cut-plane. The continuation has then be obtained by a \textit{conformal mapping of the cut-plane onto a circle}. Finally, various modifications has been introduced to optimize the summation method (for details see  \cite{rLGZJ}).\par

Further optimization of the summation technique and the additional seven-loop contributions have led to new estimates of  critical  exponents \cite{rRGZJ}. Some results are displayed in table \ref{tabCPii}.

\section{Order-dependent mapping} 

The order-dependent mapping (ODM) summation method \cite{rsezzin} is based on some  
knowledge of the analytic properties of the function that is expanded.\label{ssODMdef}
It applies both to convergent and divergent series, although it is mainly useful in
the latter case.\par
\subsection{The general method}
Let $f(z)$ be an analytic function that has the Taylor series expansion 
$$f(z)=\sum_{\ell=0} f_\ell z^\ell.$$ 
(the $=$ sign has to understood in the sense of series expansion.)\par
When the Taylor series has a finite radius of convergence, to continue the function in the whole domain of analyticity, one can map the domain onto a circle, while preserving the origin.\par
\medskip
\textit{Divergent series: the intuitive idea.}
\textit{In a case of a divergent series, one adds to the domain of analyticity a disk $|z|<r$ of variable radius $r$} and applies a similar mapping. Of course, the transformed series is still divergent. Then, one recalls the empirical rule that, for a divergent series, one is instructed to truncate the series at the term of minimal modulus, the last term giving an order of magnitude of the error. By adjusting the radius $r$ order by order, one can manage to set the minimum always just at the last calculated term. 

\begin{figure}[h]
\epsfxsize=111.8mm
\epsfysize=40.mm
\vbox{
\begin{center} 
\epsfbox{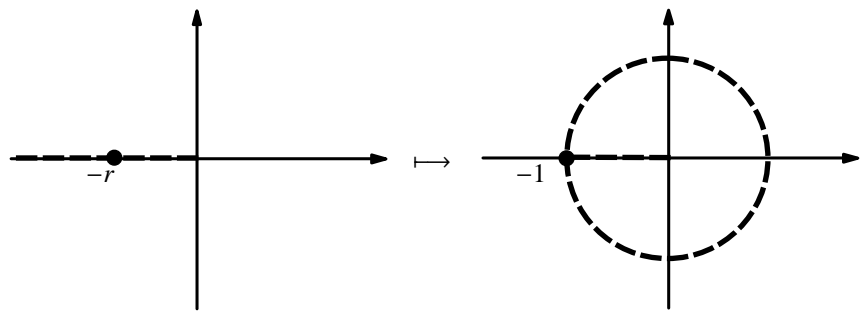}
\end{center} 
\kern-24.2mm
\moveright3mm\vbox{\hbox{\kern76.5mm$\longmapsto$}
\kern-2.5mm
\hbox{\kern33.6mm$-r$\kern52.8mm$-1$}
\kern20mm}}
\caption{Mapping $z\mapsto \lambda $: example of a function analytic in a cut-plane.}  
\label{figodmi}
\end{figure}

In what follows, we consider only functions analytic in a sector (as in the example of figure \ref{figodmi}) and mappings  \Green{$z\mapsto \lambda$} of the form 
$$z=\rho \zeta(\lambda),\quad
\zeta(\lambda)=\lambda+O\left(\lambda^2\right),$$
where \Green{$\zeta(\lambda)$} is an explicit analytic function and \Green{$\rho$} \textit{an  adjustable parameter}.\par
\textit{Although the transformed series is still divergent at  $\rho$ 
fixed, we shall verify on a few examples that, by adjusting  $\rho$  order by order} (here, we limit ourselves to Borel summable examples)  \textit{one can construct a convergent algorithm.}\par
After the transformation, \Green{$f$} is given by a Taylor series in \Green{$\lambda$} of the form
$$f\bigl( z(\lambda)\bigr)=\sum_{k=0} P_k(\rho)\lambda^k,$$
where the coefficients \Green{$P_k(\rho)$} are polynomials of degree \Green{$k$} in \Green{$\rho$}.
Since the result is formally independent of the parameter \Green{$\rho$}, the parameter can be chosen
freely.\par
The $k$-th approximant  $f^{(k)}(z)$ is constructed in the following way: one truncates the expansion at order \Green{$k$} and chooses \Green{$\rho$} as to cancel the last term. Since \Green{$P_k(\rho)$} has \Green{$k$} roots (real or complex), one chooses for \Green{$\rho$} the
largest possible root (in modulus) \Green{$\rho_k$}  for which \Green{$P_k'(\rho)$} is small. This
leads to a sequence of approximants 
$$f^{(k)}(z)=\sum_{\ell=0}^{k} P_\ell(\rho_k)\lambda^\ell(\rho_k,z) \quad  \textrm{\rm with}\quad P_k(\rho_k)=0.$$ 
In the case
of convergent series, it is expected that \Green{$\rho_k$} has a non-vanishing limit for $k\to\infty$.
By contrast, for divergent series it is expected that \Green{$\rho_k$} goes to zero
for large \Green{$k$} as
$$\rho_k=O \left( f_k ^{-1/k}\right).$$
The intuitive idea here is that \Green{$\rho_k$} corresponds to a `local' radius of convergence. \par
Since \Green{$\rho_k$} goes to zero, the function \Green{$\zeta(\lambda)$} must diverge for a finite value of \Green{$\lambda$}. Below, we choose \Green{$\lambda=1$} by convention.\par
\medskip
{\it Remark.} In the case of real functions, when the relevant zeros are complex it is
often convenient to choose minima of the polynomials \Green{$P_k$}, which satisfy
$$P'_{k}(\rho_k)=0\,,$$ 
choosing, in general, the largest zero for which \Green{$P_k$} is small.
Other mixed criteria involving a combination of $P_k$ and $P'_k$ can also be used. Indeed, the \textit{approximant is not very sensitive to the precise value of $\rho_k$, within errors}. Finally, $P_{k+1}(\rho_k)$ gives an order of magnitude of the error.

\subsection{Functions analytic in a cut-plane: Heuristic convergence analysis}

Although some rigorous convergence results have been obtained \cite{rRGKKHS}, there are not optimal. Therefore, we  give here heuristic but quantitative arguments that show the nature of the convergence of the ODM method.
Following \cite{rsezzin}, to simplify we consider a real function analytic in a cut-plane with a cut along the real negative axis (figure \ref{figodmi}) and a Cauchy representation of the form  
$$E(g)=\frac{1}{\pi}\int^{0_-}\d g'\frac{\Delta (g')}{ g'-g}\,,$$
but the generalization is simple.
Moreover, we assume that 
\begin{equation}
\Delta (g)\mathop{\propto}_{g\to 0_-} g^b\e^{A/g}  \,,\ A>0\,.\label{eDgasymptotic}
\end{equation}
The function $E(g)$ can be expanded in powers of $g$:
$$E(g)=\sum_k E_k g^k\ \textrm{with}\ E_k =\frac{1}{\pi}\int^{0_-}\frac{\d g }{ g^{k+1}}\Delta (g)\,.$$ 
The assumption (\ref{eDgasymptotic}) then implies a large order behaviour   
$$E_k\mathop{\propto}_{k\to\infty}(-A)^{-k} \Gamma(k-b)\sim (-A)^{-k} k^{-b-1}k!\,,$$
exactly of the form displayed in section \ref{LOBphiiv}.
We introduce the mapping
\begin{equation}
 g=\rho\frac{\lambda}{(1-\lambda )^\alpha }\,,\quad \alpha >1\,.\label{eODMmapping} 
 \end{equation}
The Cauchy representation then can be written as
$$E\bigl(g(\lambda )\bigr)=\frac{1}{\pi}\int^{0_-}\d  \lambda '\frac{\Delta \bigl(g(\lambda ')\bigr)}{\lambda  -\lambda '}+ R(\lambda ),$$
where $R(\lambda )$ is a sum of contributions from cuts at finite distance from the origin. 
We expand
\begin{equation}
E\bigl(g(\lambda )\bigr)=\sum_k P_k(\rho)[\lambda (g)]^k \label{eEODMlambda}
 \end{equation}
with
\begin{equation}
 P_k(\rho)=\frac{1}{\pi}\int^{0_-}\d  \lambda\, \Delta \bigl(g(\lambda  )\bigr)\lambda ^{-k-1}+ \ \hbox{finite distance contributions.}\label{eODMPkgen} 
\end{equation}
For $k\to\infty$, the factor $\lambda ^{-k}$ favours small values of $\lambda $ but for too small values of $\lambda $ the exponential decay of  $\Delta \bigl(g(\lambda  )$ takes over. Thus, $P_k(\rho_k)$ can be evaluated by the steepest descent method. With the  Ansatz that at the saddle point $\lambda< 0$ is independent of $k$ and
$$\rho_k\sim R/k\,,\ R>0\,, $$  
which implies  $g(\lambda )\to0$,  $\Delta (g)$ can be replaced by its asymptotic form (\ref{eDgasymptotic}) for $g\to0_-$. At leading order, the saddle point equation reduces to
\begin{equation}
\frac{\d}{\d \lambda }\left(\frac{A}{ R\lambda }(1-\lambda )^\alpha -\ln|\lambda |\right)=0\,.\label{eODMsaddle} 
\end{equation}
In what follows, we set   $R/A=\mu$, since this is the only parameter. 
The equation can be rewritten as
$$\mu+\frac{1}{\lambda}(1-\lambda)^{\alpha-1}\bigl((\alpha-1)\lambda+1\bigr)=0\,. $$
If the mapping (\ref{eODMmapping}) does not cancel all singularities (and this excludes the case of the integral of section \ref{ssODMintxiv}), then $P_k(\rho_k)$ cannot decrease exponentially with $k$. This implies another equation
\begin{equation}
 \frac{1}{ \lambda }(1-\lambda )^\alpha -\mu\ln|\lambda |=0\,.\label{eODMsing} 
 \end{equation}
This is indeed the region where the contribution coming from the cut at the origin and from the other finite distance singularities are comparable and where the zeros of $P_k(\rho)$ can lie.\par
Returning to the expansion (\ref{eEODMlambda}), at $g$ fixed, from the behaviour of $\rho_k$ we infer
$$1-\lambda \sim(R/kg)^{1/\alpha }\ \Rightarrow \  \lambda ^k\sim \e^{-k^{1-1/\alpha }(R/g)^{1/\alpha }}.$$ 
In a generic situation, we then expect  $P_k(\rho_k)$ to behave like
$$P_k(\rho_k)=O(\e^{Ck^{1-1/\alpha }})$$
and the domain of convergence depends on the sign of the constant $C$.
For $C>0$, the domain of convergence is
$$|g|<RC^{-\alpha }[\cos({\rm Arg}\,g/\alpha ) ]^\alpha .$$
 {\it For $\alpha >2$, this domain extends beyond the first Riemann sheet and requires analyticity of the function $E(g)$ in the corresponding domain.}\par
For $C<0$, the domain of convergence is the union of the sector $|{\rm Arg}\, g|<\pi\alpha /2$ and the domain 
$$|g|>R|C|^{-\alpha }[-\cos({\rm Arg}\,g/\alpha ) ]^\alpha .$$
Again for $\alpha >2$, this domain extends beyond the first Riemann sheet.\par
\begin{table}[t]
$$ \vbox{\offinterlineskip\tabskip=0pt\halign to \hsize
{& \vrule#\tabskip=0em plus1em & \strut\ # \ 
& \vrule#& \strut #
& \vrule#& \strut #  
& \vrule#& \strut #  
& \vrule#& \strut #  
& \vrule#& \strut #  
&\vrule#\tabskip=0pt\cr
\noalign{ \caption{} \label{tabODMsaddle} \tableskip}
\fileth
height2.0pt& \omit&& \omit&& \omit&& \omit&& \omit&& \omit&\cr
&$ \hfill{\Bfg \alpha}\hfill$&&$ \hfill\bf 3/2 \hfill$&&$ \hfill\bf 2 \hfill$&&$ \hfill \bf5/2
\hfill$&&$ \hfill\bf 3 \hfill$&&$ \hfill\bf 4 \hfill$&\cr
height2.0pt& \omit&& \omit&& \omit&& \omit&& \omit&& \omit&\cr
\fileth
height2.0pt& \omit&& \omit&& \omit&& \omit&& \omit&& \omit&\cr
&$ \hfill \mu  \hfill$&&$ \hfill4.031233504 \hfill$&&$ \hfill 4.466846120 \hfill$&&$
\hfill 4.895690188 \hfill$&&$ \hfill5.3168634291 \hfill$&&$ \hfill 6.1359656420 \hfill$&\cr 
height2.0pt& \omit&& \omit&& \omit&& \omit&& \omit&& \omit&\cr
\fileth
height2.0pt& \omit&& \omit&& \omit&& \omit&& \omit&& \omit&\cr
&$ \hfill  -\lambda\hfill$&&$ \hfill  0.2429640300 \hfill$&&$ \hfill   0.2136524524 \hfill$&&$
\hfill  0.1896450439 \hfill$&&$ \hfill  0.1699396648  \hfill$&&$ \hfill  0.14003129119 \hfill$&\cr 
height2.0pt& \omit&& \omit&& \omit&& \omit&& \omit&& \omit&\cr
\fileth
}}
$$
\end{table}
\subsection{Examples}

For $\alpha=3/2$, combining equations (\ref{eODMsaddle}) and (\ref{eODMsing}), one finds
$$\mu=4.031233504\,,\quad \lambda =-0.2429640300\,.$$

For $\alpha =2$, equation (\ref{eODMsaddle}) becomes
$$\lambda ^2-\mu \lambda -1=0 \ \Rightarrow \ \lambda =\ud(\mu-\sqrt{\mu^2+4}).$$
For $\mu=3.017759126\ldots$ one recovers the exponential rate of convergence (\ref{eODMdorate}). \par
In the case of additional singularities, with the additional equation (\ref{eODMsing}), one obtains
$$\mu=4.466846120\ldots \,,\quad \lambda =-0.2136524524\ldots .$$
To give a few other examples, again combining equations (\ref{eODMsaddle}) and (\ref{eODMsing}) one finds the results displayed in table  \ref{tabODMsaddle}.

\section{Application: The simple integral $d=0$}

For $r=1$, the integral (\ref{eODMZfiv}) in the case $d=0$ reduces to the simple integral\label{ssODMintxiv}
$$Z(g)=\frac{1}{\sqrt{2\pi}}\int\d x\,\e^{-x^2/2-g x^4/4!},$$
and the convergence of the ODM method can be studied analytically.
\subsection{The optimal mapping}

Analytic properties suggest that the optimal mapping is given by setting
$$g=\rho\frac{\lambda}{(1-\lambda)^2}\ {\rm and}\ Z(g)=(1-\lambda)^{1/2}f(\lambda).$$
Then, \Green{$f$} has an expansion of the form
$$f(\lambda)=\sum_k P_k(\rho)[\lambda(g)]^k\,.$$
Convergence can be studied analytically. First, $f$ has the representation
$$f(\lambda)=\frac{1}{\sqrt{2\pi}}\int\d s\,\e^{-s^2/2+\lambda (s^2/2-\rho s^4/24)}
=P_k(\rho)[\lambda(g)]^k$$
with
$$P_k(\rho)=\frac{1}{ k!}\frac{1}{\sqrt{2\pi}}\int\d s\,\e^{-s^2/2}(s^2/2-\rho s^4/24)^k .$$
Setting
$$s^2/2=kt\,,\quad \rho=R/k\,,$$
one can rewrite the expression as
$$P_k(\rho)= \frac{k^{k+1/2}}{ k!}\sqrt{\frac{1}{\pi}}\int\frac{\d t}{\sqrt{t}}\e^{-kt} (t-Rt^2/6)^k.$$
For $k\to\infty$, the integral can be evaluated by the steepest descent method. The saddle point equation is
$$ Rt^2/6-(1+R/3)t+1=0  $$
and, thus,
$$t=\frac{3}{ R}\left(1+R/3\pm\sqrt{1+R^2/9}\right).$$
For $k$ odd, the zero corresponds to a cancellation between the two saddle points. This yields the equation
$$\e^{2\sqrt{R^2+9}/R}=\frac{\sqrt{R^2+9}+R}{ \sqrt{R^2+9}-R}\ \Leftrightarrow \ \e^{\sqrt{R^2+9}/R}= \frac{1}{3} \left(\sqrt{R^2+9}+R\right) .$$ 
 As expected, one finds
 $$\rho_k\sim\frac{R }{ k}\ {\rm with} \ R=4.526638689\ldots \ (R/A=3.017759126\ldots) .$$  
The minimum is given by one of the saddle points
\begin{equation}
P_k(\rho)\propto \e^{-3k/R }= (0.5154353381\ldots)^k.\label{eODMdorate}
\end{equation}
At  $g$ fixed,  $\lambda$  converges to  $1$. More precisely,
 $$\lambda=1-\sqrt{R/kg}+O(1/k)\ \Rightarrow \ \lambda^k\sim \e^{-\sqrt{Rk/g}}.$$  
 {\it The approximants converge geometrically on the entire Riemann surface}, a situation possible only because the function  $Z(g)$  has no other singularity at finite distance.\par

\begin{table}[t]
$$ \vbox{\offinterlineskip\tabskip=0pt\halign to \hsize
{& \vrule#\tabskip=0em plus1em & \strut\ # \ 
& \vrule#& \strut #
& \vrule#& \strut #  
& \vrule#& \strut #  
& \vrule#& \strut #  
& \vrule#& \strut #  
& \vrule#& \strut # 
&\vrule#\tabskip=0pt\cr
\noalign{\caption{}  \label{tabODMii}\tableskip}
\noalign{\it ODM for the integral $d=0$ for $g\to\infty$: $Z(g)\sim g^{-1/4}(1/2)*24^{1/4}*\sqrt{\pi}/\Gamma(3/4)$. We define $[Z(g)  g^{1/4}]_{\rm exact}- [Z(g)  g^{1/4}]_{\rm ODM}=\delta$ \tableskip} 
\fileth
height2.0pt& \omit&& \omit&& \omit&& \omit&& \omit&& \omit&& \omit&\cr
&$ \hfill\bf k \hfill$&&$ \hfill\bf 5 \hfill$&&$ \hfill\bf 10 \hfill$&&$ \hfill \bf15
\hfill$&&$ \hfill\bf 20 \hfill$&&$ \hfill\bf 25 \hfill$&&$ \hfill\bf 30 \hfill$&\cr
height2.0pt& \omit&& \omit&& \omit&& \omit&& \omit&& \omit&& \omit&\cr
\fileth
height2.0pt& \omit&& \omit&& \omit&& \omit&& \omit&& \omit&& \omit&\cr
&$ \hfill 1/\rho  \hfill$&&$ \hfill 1.131726  \hfill$&&$ \hfill 2.35036 \hfill$&&$
\hfill  3.34050 \hfill$&&$ \hfill  4.5594 \hfill$&&$ \hfill 5.5495 \hfill$&&$
\hfill 6.8614 \hfill$&\cr 
height2.0pt& \omit&& \omit&& \omit&& \omit&& \omit&& \omit&& \omit&\cr
\fileth
height2.0pt& \omit&& \omit&& \omit&& \omit&& \omit&& \omit&& \omit&\cr
&$ \hfill -\delta \hfill$&&$ \hfill  5.7*10^{-3} \hfill$&&$ \hfill   2.5*10^{-5} \hfill$&&$
\hfill   3.7*10^{-6} \hfill$&&$ \hfill  2.2*10^{-8}  \hfill$&&$ \hfill  3.4*10^{-9} \hfill$&&$
\hfill5.1*10^{-11} \hfill$&\cr 
height2.0pt& \omit&& \omit&& \omit&& \omit&& \omit&& \omit&& \omit&\cr
\fileth
height2.0pt& \omit&& \omit&& \omit&& \omit&& \omit&& \omit&& \omit&\cr
&$ \hfill \ln|\delta | \hfill$&&$ \hfill -5.1578 \hfill$&&$ \hfill  -10.5921
\hfill$&&$ \hfill  -12.5008\hfill$&&$ \hfill -17.5923 \hfill$&&$ \hfill -19.4855
\hfill$&&$ \hfill  -23.6818 \hfill$&\cr 
height2.0pt& \omit&& \omit&& \omit&& \omit&& \omit&& \omit&& \omit&\cr
\fileth
height2.0pt& \omit&& \omit&& \omit&& \omit&& \omit&& \omit&& \omit&\cr
&$\bf \hfill k \hfill$&&$ \hfill\bf 35 \hfill$&&$ \hfill\bf 40 \hfill$&&$ \hfill\bf 45
\hfill$&&$ \hfill\bf 50 \hfill$&&$ \hfill\bf 55 \hfill$&&$ \hfill\bf 60 \hfill$&\cr
height2.0pt& \omit&& \omit&& \omit&& \omit&& \omit&& \omit&& \omit&\cr
\fileth
height2.0pt& \omit&& \omit&& \omit&& \omit&& \omit&& \omit&& \omit&\cr
&$ \hfill 1/\rho  \hfill$&&$ \hfill 7.7586  \hfill$&&$ \hfill 8.9778 \hfill$&&$
\hfill  9.9678 \hfill$&&$ \hfill  11.1869 \hfill$&&$ \hfill12.1769 \hfill$&&$
\hfill 13.3958 \hfill$&\cr 
height2.0pt& \omit&& \omit&& \omit&& \omit&& \omit&& \omit&& \omit&\cr
\fileth
height2.0pt& \omit&& \omit&& \omit&& \omit&& \omit&& \omit&& \omit&\cr
&$ \hfill -\delta  \hfill$&&$ \hfill   3.5*10^{-12} \hfill$&&$ \hfill   2.5*10^{-14} \hfill$&&$
\hfill   3.9*10^{-15} \hfill$&&$ \hfill 2.9*10^{-17}  \hfill$&&$ \hfill  4.5*10^{-18} \hfill$&&$
\hfill3.4*10^{-20} \hfill$&\cr 
height2.0pt& \omit&& \omit&& \omit&& \omit&& \omit&& \omit&& \omit&\cr
\fileth
height2.0pt& \omit&& \omit&& \omit&& \omit&& \omit&& \omit&& \omit&\cr
&$ \hfill \ln|\delta | \hfill$&&$ \hfill -26.3535 \hfill$&&$ \hfill  -31.2859
\hfill$&&$ \hfill  -33.1625\hfill$&&$ \hfill -38.0643 \hfill$&&$ \hfill -39.9364
\hfill$&&$ \hfill  -44.8208 \hfill$&\cr 
height2.0pt& \omit&& \omit&& \omit&& \omit&& \omit&& \omit&& \omit&\cr
\fileth
}}
$$
\end{table}
\smallskip
{\it Numerical verifications.} With about 60 terms, the slope is found to be $1/k\rho_k\approx 0.2209$ in agreement with the prediction $1/R=0.2209$ (once even--odd order oscillations are taken into account). The logarithm of the error has a slope $0.696/0.685$ to be compared with the prediction $3/R=0.66$ (see table \ref{tabODMii}).
\subsection {An alternative mapping}

Another mapping that also regularizes the point at infinity is
$$g=\rho\frac{\lambda}{(1-\lambda)^4}. $$
Numerical results for $g=5$ are displayed in table \ref{tabODMiii}. 
\begin{table}[t]
$$ \vbox{\offinterlineskip\tabskip=0pt\halign to \hsize
{& \vrule#\tabskip=0em plus1em & \strut\ # \ 
& \vrule#& \strut #
& \vrule#& \strut #  
& \vrule#& \strut #  
& \vrule#& \strut #  
& \vrule#& \strut #  
& \vrule#& \strut # 
&\vrule#\tabskip=0pt\cr
\noalign{  \caption{} \label{tabODMiii}\tableskip}
\noalign{ \it  ODM for the integral $d=0$ for $g=5$:   we define $[Z(g)   ]_{\rm exact}- [Z(g)   ]_{\rm ODM}=\delta$ \tableskip} 
\fileth
height2.0pt& \omit&& \omit&& \omit&& \omit&& \omit&& \omit&& \omit&\cr
&$ \hfill\bf k \hfill$&&$ \hfill\bf 5 \hfill$&&$ \hfill\bf 10 \hfill$&&$ \hfill \bf15
\hfill$&&$ \hfill\bf 20 \hfill$&&$ \hfill\bf 25 \hfill$&&$ \hfill\bf 30 \hfill$&\cr
height2.0pt& \omit&& \omit&& \omit&& \omit&& \omit&& \omit&& \omit&\cr
\fileth
height2.0pt& \omit&& \omit&& \omit&& \omit&& \omit&& \omit&& \omit&\cr
&$ \hfill 1/\rho  \hfill$&&$ \hfill 0.5918 \hfill$&&$ \hfill 1.0297 \hfill$&&$
\hfill  1.5627\hfill$&&$ \hfill  2.0779 \hfill$&&$ \hfill 2.5865 \hfill$&&$
\hfill 3.1376 \hfill$&\cr 
height2.0pt& \omit&& \omit&& \omit&& \omit&& \omit&& \omit&& \omit&\cr
\fileth
height2.0pt& \omit&& \omit&& \omit&& \omit&& \omit&& \omit&& \omit&\cr
&$ \hfill |\delta| \hfill$&&$ \hfill  1.1*10^{-3} \hfill$&&$ \hfill   3.7*10^{-5} \hfill$&&$
\hfill   1.7*10^{-6} \hfill$&&$ \hfill  9.2*10^{-7}  \hfill$&&$ \hfill  1.1*10^{-7} \hfill$&&$
\hfill3.5*10^{-9}\hfill$&\cr 
height2.0pt& \omit&& \omit&& \omit&& \omit&& \omit&& \omit&& \omit&\cr
\fileth
height2.0pt& \omit&& \omit&& \omit&& \omit&& \omit&& \omit&& \omit&\cr
&$ \hfill \ln|\delta | \hfill$&&$ \hfill -6.7454 \hfill$&&$ \hfill  -10.2069
\hfill$&&$ \hfill  -13.2837\hfill$&&$ \hfill -13.8898 \hfill$&&$ \hfill -16.0103
\hfill$&&$ \hfill  -19.4614 \hfill$&\cr 
height2.0pt& \omit&& \omit&& \omit&& \omit&& \omit&& \omit&& \omit&\cr
\fileth
height2.0pt& \omit&& \omit&& \omit&& \omit&& \omit&& \omit&& \omit&\cr
&$\bf \hfill k \hfill$&&$ \hfill\bf 35 \hfill$&&$ \hfill\bf 40 \hfill$&&$ \hfill\bf 45
\hfill$&&$ \hfill\bf 50 \hfill$&&$ \hfill\bf 55 \hfill$&&$ \hfill\bf 60 \hfill$&\cr
height2.0pt& \omit&& \omit&& \omit&& \omit&& \omit&& \omit&& \omit&\cr
\fileth
height2.0pt& \omit&& \omit&& \omit&& \omit&& \omit&& \omit&& \omit&\cr
&$ \hfill 1/\rho  \hfill$&&$ \hfill 3.6167  \hfill$&&$ \hfill4.1877  \hfill$&&$
\hfill 4.6557  \hfill$&&$ \hfill 5.3021  \hfill$&&$ \hfill5.6959  \hfill$&&$
\hfill6.2458 \hfill$&\cr 
height2.0pt& \omit&& \omit&& \omit&& \omit&& \omit&& \omit&& \omit&\cr
\fileth
height2.0pt& \omit&& \omit&& \omit&& \omit&& \omit&& \omit&& \omit&\cr
&$ \hfill -\delta  \hfill$&&$ \hfill   3.4*10^{-9} \hfill$&&$ \hfill  8.0*10^{-10}   \hfill$&&$
\hfill 1.0*10^{-10}   \hfill$&&$ \hfill 2.9*10^{-11}   \hfill$&&$ \hfill    1.2*10^{-11}\hfill$&&$
\hfill2.4*10^{-12}  \hfill$&\cr 
height2.0pt& \omit&& \omit&& \omit&& \omit&& \omit&& \omit&& \omit&\cr
\fileth
height2.0pt& \omit&& \omit&& \omit&& \omit&& \omit&& \omit&& \omit&\cr
&$ \hfill \ln|\delta | \hfill$&&$ \hfill -19.4706 \hfill$&&$ \hfill  -20.9453 
\hfill$&&$ \hfill -22.9796  \hfill$&&$ \hfill -24.2450  \hfill$&&$ \hfill-25.0907  
\hfill$&&$ \hfill -26.7433   \hfill$&\cr 
height2.0pt& \omit&& \omit&& \omit&& \omit&& \omit&& \omit&& \omit&\cr
\fileth
}}
$$
\end{table}
Finally, for $k=65$, $70$,
$$1/\rho_k=6.7479\,,\ 7.5724\,,\quad \ln|\delta|=-30.3657\,,\ -28.9505 $$
On the average between $k=5$ and $k=70$,
$$\rho_k\sim R/k\ {\rm with}\ R=9.75\,. $$
The results displayed in table  \ref{tabODMsaddle} lead to the prediction
$$R=9.2039 \,.$$ 
The error is about
$$\delta=-1.59\,\frac{k^{3/4}}{ g^{1/4}} $$
and $1.59$ has to be compared with the expected asymptotic value $1.74$ if one assumes convergence for all $g>0$.
\section{The quartic anharmonic oscillator: $d=1$}

For $r=1$, the path integral corresponds to the quantum Hamiltonian
$$H=\ud p^2+\ud x^2+\frac{g}{4!} x^4.$$
The eigenvalues $E$ of $H$ are given by the solution of the time-independent Schrödinger equation
$$-\ud\psi''(x)+\left(\ud x^2+\frac{g}{4!} x^4\right)\psi(x)=E\psi(x),$$
where $\psi(x)$ is a square-integrable function.\par
As an example, we consider the perturbative expansion of the lowest eigenvalue, the ground state energy. 
Variational arguments and scaling suggest the mapping
\begin{equation}
g=\rho\frac{\lambda}{(1-\lambda)^{3/2}}\,,\quad E=\frac{\mathcal{ E}}{
(1-\lambda)^{1/2}}\,.\label{emapanh} 
\end{equation}
Then,
$$\mathcal{ E}=\sum_k P_k(\rho)[\lambda(g)]^k.$$
Large order behaviour (section \ref{LOBphiiv}) and a steepest descent evaluation (table \ref{tabODMsaddle}) lead to the prediction
$$\rho_k\sim R/k\ {\rm with}\ R=\mu A=32.25\ldots.$$
Then, $\lambda$ converges to \Green{$1$} as 
$$\lambda=1-\left(\frac{R}{ kg}\right)^{2/3}+O( k^{-4/3})\ \Rightarrow\ \lambda^k\sim\e^{-R^{2/3} k^{1/3}/g^{2/3}}\ {\rm with}\  R^{2/3}=10.131\ldots.$$  
 An unbiased fit of the numerical data for $k\le 60$ yields results within 10\% of the predicted values.
Finally, a fit of the relative error for $g\to\infty$ yields \cite{rsezzin}
$$P_{k+1}(\rho_k)\propto \e^{-9.6 k^{1/3}}.$$ 
The relative error at order \Green{$k$} is thus of order \Green{$\e^{-k^{1/3}(9.6+R^{2/3}g^{-2/3})}$}.
One finds convergence for
 $$0.95+|g|^{-2/3}\cos\left(\textstyle{\frac{2}{3}}{\rm Arg}\,g\right)>0\,.$$ 
The corresponding domain contains a section of the first Riemann sheet and extends to the second Riemann sheet for  $|g|$ large enough. 
\section{$\phi^4$ field theory in $d=3$ dimensions}

In \cite{rsezzin}, the ODM method has been applied on functions of the initial parameter $g$ of the action (\ref{eODMfiv}) rather the renormalized parameter $g_{\rm r}$ introduced in section \ref{ssBorelmapping}.
Then the point of physical interest is $g\to\infty$, which corresponds to the zero $\g^*$ of the $\beta$-function (\ref{ebetafiviiiIsing}). Due to UV divergences, a needed regularization and renormalization,  scaling arguments are no longer applicable to determine an appropriate mapping. The relation (\ref{erbetadef}) between initial and renormalized parameter shows that,
for $g\to\infty$, physical observables have an expansion in powers of
$g^{-\omega}$, where the exponent  $\omega=\tilde\beta'(\g^*)$.
This then suggests the mapping
\begin{equation}
g=\rho\frac{\lambda }{(1-\lambda)^{1/\omega}}, 
\end{equation} 
but the difficulty is that $\omega$ has to be inferred from the series (\ref{ebetafiviiiIsing}) itself. The results obtained in this way \cite{rsezzin} are consistent with those obtained in \cite{rRGZJ}  ($\omega=0.80(1)$ from Borel transformation and mapping), but   empirical errors are more difficult to assess. Also the expected rate of convergence is of order
\Green{$\e^{-{\rm const.} k^{1-\omega}}=\e^{-{\rm const.}k^{0.2}}$}, which is rather slow (see table \ref{tableODM3D}). See reference  \cite{rsezzin} for details. Finally, the information about the large order behaviour cannot easily be incorporated.\par

\begin{table}[h]
$$  \offinterlineskip\tabskip=0pt\halign to \hsize
{& \vrule#\tabskip=0em plus1em & \strut\ # \ 
& \vrule#& \strut #
& \vrule#& \strut #  
& \vrule#& \strut #  
& \vrule#& \strut #  
& \vrule#& \strut #   
&\vrule#\tabskip=0pt\cr
\noalign{\caption{}\label{tableODM3D} \tableskip}
\noalign{\noindent\it Series for the exponent \Green{$\omega$} summed by ODM in the \Green{$ \phi^4_3 $} field
theory with \Green{$\omega_{\rm in}=0.79$} and \Green{$\d\omega_{\rm cal.}/\d\omega_{\rm in}=-0.6$}.\tableskip} 
\fileth
height2.0pt& \omit&& \omit&& \omit&& \omit&& \omit&& \omit&\cr
&$ \hfill k \hfill$&&$ \hfill 2 \hfill$&&$ \hfill 3 \hfill$&&$ \hfill 4
\hfill$&&$ \hfill 5 \hfill$&&$ \hfill 6 \hfill$&\cr
height2.0pt& \omit&& \omit&& \omit&& \omit&& \omit&& \omit&\cr
\fileth
height2.0pt& \omit&& \omit&& \omit&& \omit&& \omit&& \omit&\cr
&$ \hfill\omega_k \hfill$&&$ \hfill 0.552  \hfill$&&$ \hfill 0.754 \hfill$&&$
\hfill 0.711 \hfill$&&$ \hfill  0.767 \hfill$&&$ \hfill 0.759 \hfill$&\cr 
height2.0pt& \omit&& \omit&& \omit&& \omit&& \omit&& \omit&\cr
\fileth}
$$
\end{table}
Here, to illustrate the flexibility of the method, we work directly with functions of $\g$. We also take into account the covariance of the $\beta$-functions under a change of parametrization:
\begin{equation}
 \beta_1(g_1)=\frac{\d g_1}{\d g_2}\beta_2(g_2).\label{ebetacov}
 \end{equation}
This transformation law is such that the derivative of the $\beta$-function at a zero (a fixed point), which is a physical observable, remains unchanged.\par
Equation (\ref{ebetacov}) suggests a mapping of the form
$$\g=\rho\left(\frac{1}{(1-\lambda)^\alpha}-1\right),$$
with a suitable choice of the parameter $\alpha$, since unlike a mapping of the form (\ref{eODMmapping}), it introduces no new singularity. We thus set
$$\beta_\lambda(\lambda)=\frac{(1-\lambda)^{\alpha+1}}{\alpha\rho}\tilde\beta\bigl(\g(\lambda)\bigr).$$
A few trials, without trying to optimize, suggest  the value $\alpha=3/2$ and this is the value we have adopted. The results for the zero $\tilde g^*$, for comparison with the method outlined in section \ref{ssBorelmapping}, and the exponent $\omega=\beta'_\lambda(\lambda^*)$ are given in table \ref{tableODM3Db}. The order of magnitude of the errors can only be estimated by the sensitivity to the precise choice of the parameter $\rho$ (zero of last term or its derivative, for example). The indications are $\Delta\g^*\approx\Delta\omega\approx 0.006$.
In the case of complex zeros, we have given only the real part in the tables.
\begin{table}[h]
$$  \offinterlineskip\tabskip=0pt\halign to \hsize
{& \vrule#\tabskip=0em plus1em & \strut\ # \ 
& \vrule#& \strut #
& \vrule#& \strut #  
& \vrule#& \strut #  
& \vrule#& \strut #  
& \vrule#& \strut #   
&\vrule#\tabskip=0pt\cr
\noalign{\caption{}\label{tableODM3Db} \tableskip}
\noalign{\noindent\it Results for the zero $\g^*$ of the $\beta$-function and the exponent \Green{$\omega$} summed by ODM in the  $ \phi^4_3 $  field
theory. With the method of section \ref{ssBorelmapping} (\cite{rRGZJ}, one finds $\g^*=1.411 \pm 0.004$ and $\omega=0.799 \pm 0.011$.\tableskip} 
\fileth
height2.0pt& \omit&& \omit&& \omit&& \omit&& \omit&& \omit&\cr
&$ \hfill k \hfill$&&$ \hfill 3 \hfill$&&$ \hfill 4 \hfill$&&$ \hfill 5
\hfill$&&$ \hfill 6 \hfill$&&$ \hfill 7 \hfill$&\cr
height2.0pt& \omit&& \omit&& \omit&& \omit&& \omit&& \omit&\cr
\fileth
height2.0pt& \omit&& \omit&& \omit&& \omit&& \omit&& \omit&\cr
&$ \hfill \tilde g^* \hfill$&&$ \hfill 1.09871  \hfill$&&$ \hfill 1.39330 \hfill$&&$
\hfill 1.41771 \hfill$&&$ \hfill  1.41737 \hfill$&&$ \hfill 1.41744 \hfill$&\cr 
height2.0pt& \omit&& \omit&& \omit&& \omit&& \omit&& \omit&\cr
\fileth
height2.0pt& \omit&& \omit&& \omit&& \omit&& \omit&& \omit&\cr
&$ \hfill\omega  \hfill$&&$ \hfill 1  \hfill$&&$ \hfill 0.7984 \hfill$&&$
\hfill 0.7804 \hfill$&&$ \hfill  0.7806 \hfill$&&$ \hfill 0.7807 \hfill$&\cr 
height2.0pt& \omit&& \omit&& \omit&& \omit&& \omit&& \omit&\cr
\fileth
}
$$
\end{table}\par
Other critical exponents are obtained from the two RG functions 
\begin{eqnarray*}
\gamma^{-1}(\g)&=&1-\textstyle{\frac{1}{6}}  \,\g  +\textstyle{\frac{1}{27}}\, \g  ^{2}- 0.0230696212  \,\g ^{3}+ 0.0198868202  \,\g ^{4}- 0.0224595215 \,\g^{5} \\&&\mbox{}+0.0303679053  \mbox{}\,\g^{6}-0.046877951  \,g^7+O(g^8), \\
\eta(\g)&=& 0.0109739368  \,\g^{2}+ 0.0009142222 \,\g^{3}+ 0.0017962228  {g}^{4}- 0.0006537035  \,{\g}^{5}\\ &&\mbox{}+ 0.0012749100 \,{\g}^{6}- 0.001697694 \,{\g}^{7}+O(g^8),
\end{eqnarray*}
by setting $\g=\g^*$.
We have also summed the series for $\gamma^{-1}(\g)$ and $\eta(\g)/\g^2$, and independently the series for the function $\nu^{-1}(\g)$ although it is related to the $\gamma$ and $\eta$ by $\gamma(\g)=\nu(\g)(2-\eta(\g))$. A verification of this relation after summation gives an indication about the errors. The results, displayed in table \ref{tableODM3Dc}, can be compared with the results of \cite{rRGZJ}:
$$\gamma=1.2396 \pm 0.0013\,,\quad\nu=0.6304 \pm 0.0013\,,\quad \eta=0.0335 \pm 0.0025\,.$$
One notices a reasonable consistency.
\begin{table}[h]
$$  \offinterlineskip\tabskip=0pt\halign to \hsize
{& \vrule#\tabskip=0em plus1em & \strut\ # \ 
& \vrule#& \strut #
& \vrule#& \strut #  
& \vrule#& \strut #  
& \vrule#& \strut #  
& \vrule#& \strut #   
&\vrule#\tabskip=0pt\cr
\noalign{\caption{}\label{tableODM3Dc} \tableskip}
\noalign{ \it Critical exponents $\gamma=\gamma(\g^*)$, $\nu=\nu(\g^*)$ and $\eta=\eta(\g^*)$ for $\g^*=1.411$.\tableskip} 
\fileth
height2.0pt& \omit&& \omit&& \omit&& \omit&& \omit&& \omit&\cr
&$ \hfill k \hfill$&&$ \hfill 3 \hfill$&&$ \hfill 4 \hfill$&&$ \hfill 5
\hfill$&&$ \hfill 6 \hfill$&&$ \hfill 7 \hfill$&\cr
height2.0pt& \omit&& \omit&& \omit&& \omit&& \omit&& \omit&\cr
\fileth
height2.0pt& \omit&& \omit&& \omit&& \omit&& \omit&& \omit&\cr
&$ \hfill  \gamma \hfill$&&$ \hfill 1.23717  \hfill$&&$ \hfill 1.23486 \hfill$&&$
\hfill 1.23845 \hfill$&&$ \hfill  1.23820 \hfill$&&$ \hfill 1.23923 \hfill$&\cr 
height2.0pt& \omit&& \omit&& \omit&& \omit&& \omit&& \omit&\cr
\fileth
height2.0pt& \omit&& \omit&& \omit&& \omit&& \omit&& \omit&\cr
&$ \hfill \nu  \hfill$&&$ \hfill 0.62521  \hfill$&&$ \hfill 0.62486 \hfill$&&$
\hfill 0.62746 \hfill$&&$ \hfill  0.62771 \hfill$&&$ \hfill 0.62865 \hfill$&\cr 
height2.0pt& \omit&& \omit&& \omit&& \omit&& \omit&& \omit&\cr
\fileth
height2.0pt& \omit&& \omit&& \omit&& \omit&& \omit&& \omit&\cr
&$ \hfill \eta  \hfill$&&$ \hfill    \hfill$&&$ \hfill0.0290   \hfill$&&$
\hfill 0.0289\hfill$&&$ \hfill  0.0297 \hfill$&&$ \hfill 0.0306 \hfill$&\cr 
height2.0pt& \omit&& \omit&& \omit&& \omit&& \omit&& \omit&\cr
\fileth
}
$$
\end{table}
\section{Convergence proofs and physics applications}

Since  the ODM summation method has been proposed, convergence  has been proved in specific cases \cite{rRGKKHS}, however, to our knowledge, systematic mathematical investigations are still lacking and would be most welcome.\par
Notice that in \cite{rWECas}  a special case of the ODM method has also been proposed. Moreover, methods known as \textit{linear or scaled delta expansion}, or optimized perturbation theory, introduced later, often reduce  to special cases of the ODM method (see, for example, \cite{rDuJo},\cite{rBDJ}).\par
 The ODM  method has also found a number of useful applications in physics (sometimes under different names). In \cite{rLGZJHyd}, the problem of the hydrogen atom in strong magnetic fields has been considered. The summation of the weak field series expansion has led to precise determinations of the ground state energy for very strong fields. In \cite{rHKWJ}, the location of the Bender--Wu singularities of the quartic anharmonic oscillator has been determined numerically (for the anharmonic oscillator see also \cite{rBBPGAN}).
In \cite{rRGZJeq}, the equation of state of physical systems belonging to the Ising model universality class has been determined numerically by a combination of Borel transformation and mapping and the ODM summation method.  Another application has been the Bose--Einstein condensation problem \cite{rSCPRS}, \cite{rJLKMPRR} (for similar problems see also \cite{rEBER}). Finally, let us also mention the application to the Gross--Neveu model \cite{rKPR}, \cite{rKPRS}.

{\par\bigskip

}
\end{document}